\documentclass[%
 reprint,
 superscriptaddress,
 amsmath,amssymb,
 aps,
]{revtex4-2}

\usepackage{graphicx}
\usepackage{dcolumn}
\usepackage{bm}
\usepackage{hyperref}

\begin{document}

\title{Stochastic dynamics without detailed balance condition connecting simple gradient method and Hamiltonian Monte Carlo}

\author{Akihisa Ichiki}
\email{ichiki@chem.material.nagoya-u.ac.jp}
\affiliation{Institutes of Innovation for Future Society, Nagoya University, Furo-cho, Chikusa-ku, Nagoya 464-8603, Japan}
\author{Masayuki Ohzeki}
\email{masayuki.ohzeki.a4@tohoku.ac.jp}
\affiliation{Graduate School of Information Sciences, Tohoku University, Sendai 980-8579, Japan}
\affiliation{Institute of Innovative Research, Tokyo Institute of Technology, Oh-okayama, Meguro-ku, Tokyo 152-8550, Japan}
\affiliation{Sigma-i, Co. Ltd., Konan, Minato-ku, Tokyo 108-0075, Japan}
\date{\today}

\begin{abstract}
Sampling occupies an important position in theories of various scientific fields, and Markov chain Monte Carlo~(MCMC) provides the most common technique of sampling. In the progress of MCMC, a huge number of studies have aimed the acceleration of convergence to the target distribution. Hamiltonian Monte Carlo~(HMC) is such a variant of MCMC. In the recent development of MCMC, another approach based on the violation of the detailed balance condition has attracted much attention. Historically, these two approaches have been proposed independently, and their relationship has not been clearly understood. In this paper, the two approaches are seamlessly understood in the framework of generalized Monte Carlo method that violates the detailed balance condition. Furthermore we propose an efficient Monte Carlo method based on our framework. 
\end{abstract}

\maketitle

\section{introduction}
Recently, sampling techniques have become of increasing importance in various fields of science and engineering. The sampling methods have been developed to numerically examine the equilibrium behaviors of complex systems such as macromolecules like proteins~\cite{10.1006/jcph.1999.6231, Mitsutake2001}, spin glasses~\cite{PhysRevB.32.7384}, and glass transitions~\cite{PhysRevE.61.5473}. In addition to these traditional applications, with the background of the recent development of machine learning, sampling has become widely used for various purposes such as model training and its evaluation, and stochastic inference~\cite{Andrieu2003, 10.5555/971143}.

The most common technique for sampling is provided by the Markov chain Monte Carlo~(MCMC) method. MCMC is required to quickly sample random variables that follow an arbitrary target distribution starting from a given initial state. Since Metropolis~{\it et~al.} successfully introduced MCMC to investigate complex systems~\cite{Metropolis1953}, many variants have been proposed to accelerate the convergence to the target distribution. The speed-up techniques have been constructed mainly based on two concepts. One is called the extended ensemble method~\cite{IBA2001}. In the extended ensemble method, the state space is extended by introducing auxiliary variables, and the convergence is accelerated by a proposal of a path in higher dimension allowing a rapid transition to the target distribution. The techniques of extended ensemble are roughly categorized into three groups: the exchange Monte Carlo~\cite{Hukushima1996}, the simulated tempering~\cite{Marinari_1992, Lyubartsev1992}, and the multicanonical method~\cite{PhysRevLett.68.9} with the help of the Wang-Landau algorithm~\cite{PhysRevLett.86.2050}. Hamiltonian Monte Carlo~(HMC), which introduces momenta as auxiliary variables~\cite{Duane1987}, is also classified as an extended ensemble method. The alternative concept for acceleration is based on an efficient proposal of candidates for the updated state. Such efficient candidates are generated via the concept of the coarse-graining. The Swendensen-Wang algorithm~\cite{PhysRevLett.58.86} makes efficient state updates by using clusters of spins in the Ising model. This algorithm was later extended by Wolff to $XY$ model~\cite{PhysRevLett.62.361}, and is now extended to be applied to an arbitrary target distribution~\cite{Barbu2005}.

Recently, in addition to the above mentioned two concepts for acceleration, the possibility of detailed balance violation has been intensively investigated~\cite{PhysRevLett.105.120603, Turitsyn2011, FERNANDES20111856, doi:10.7566/JPSJ.82.064003, doi:10.7566/JPSJ.82.064003}. Conventional acceleration algorithms have been developed within the range of the detailed balance condition. However, it has been shown that the violation of the detailed balance accelerates convergence to the target distributions~\cite{PhysRevE.88.020101}. Based on this result, Ohzeki and Ichiki proposed a systematic construction of detailed balance-violating dynamics that converges to any target distribution in a continuous system~\cite{PhysRevE.92.012105}. The Ohzeki-Ichiki method duplicates the original system and introduces a probability current between the two systems. The driving force producing the probability current causes the rotational evolution of state in the duplicated state space. This is similar to the symplectic behavior of the Hamiltonian dynamics. In this paper, the Ohzeki-Ichiki method will be generalized, and it will be explained that the generalized Ohzeki-Ichiki method is indeed seamlessly connected to the Hamiltonian dynamics.

The generalized Ohzeki-Ichiki method provides a family of dynamics including the gradient method and the HMC. To show this fact, after reviewing the gradient method in section~\ref{sec:grad}, the HMC in section~\ref{sec:HMC}, and the Ohzeki-Ichiki method in section~\ref{sec:OI}, respectively, we will see that the generalized Ohzeki-Ichiki method contains the gradient method and the HMC as specific limits. In section~\ref{sec:hybrid}, the generalized Ohzeki-Ichiki method is numerically compared to other methods with respect to the speed of convergence to the target distribution. Section~\ref{sec:summary} is devoted to a summary and discussion.

\section{gradient method}\label{sec:grad}
The simplest dynamics converging to the target distribution is given by a gradient method. The gradient method satisfies the so-called detailed balance condition. Physically, the dynamics with the detailed balance condition is relaxed to a steady state in which no macroscopic heat is generated. Such a special steady state is called an equilibrium state. By the gradient method, the Gibbs distribution 
\begin{eqnarray}
\pi(x) = \exp\left[-U(x)/T\right] / Z\label{Gibbs}
\end{eqnarray}
with a partition function $Z$ is achieved with the balance between the energy gradient and the diffusion due to noise. The following dynamics gives the simplest gradient method in which the $N$-dimensional continuous state $x$ converges to the Gibbs distribution: 
\begin{eqnarray}
dx_i(t) = -\dfrac{\partial U}{\partial x_i}dt + \sqrt{2T}dW_i(t)\,,\label{gradLangevin}
\end{eqnarray}
where, $dx_i$ is the displacement of $x_i$ during an infinitesimal time $dt$, and $U(x)$ and $T$ correspond to the potential and temperature, respectively. $W_i(t)$ is a standard Wiener process that satisfies 
\begin{eqnarray}
\left\langle dW_i(t) \right\rangle &=& 0\,,\\
\left\langle dW_i(t) dW_j(t’)\right\rangle &=& \delta_{ij}\delta(t - t’)dt\,,
\end{eqnarray}
where $\delta_{ij}$ and $\delta(t)$ denote Kronecker and Dirac delta functions, respectively, and $\left\langle\cdot\right\rangle$ represents an expectation. The Fokker-Planck equation corresponding to the Langevin equation~(\ref{gradLangevin}) is given as 
\begin{eqnarray}
\dfrac{\partial P(x, t)}{\partial t} = -\displaystyle\sum_i \dfrac{\partial}{\partial x_i}\left[ -\dfrac{\partial U(x)}{\partial x_i} – T\dfrac{\partial}{\partial x_i}\right] P(x, t)\,.\label{gradFPE}
\end{eqnarray}
It is straightforwardly confirmed that the Gibbs distribution~(\ref{Gibbs}) is the steady solution satisfying the Fokker-Planck equation~(\ref{gradFPE}). 

It is guaranteed by the H-theorem that the dynamics~(\ref{gradLangevin}) converges to a unique steady distribution~(\ref{Gibbs}) as an equilibrium distribution regardless of an initial condition. Therefore, the target Gibbs distribution can be obtained by providing $U(x)$ and $T$ in the simple gradient dynamics~(\ref{gradLangevin}). However, since the simple gradient method updates the state along the gradient of the potential $U$, the update becomes inefficient when the state is trapped in a local minimum of the potential, where the gradient vanishes. To escape from such a local minimum, noise is exploited in MCMC algorithms. However, if the potential around the local minimum is steep, it takes a long time to escape from the local minimum. In the history of MCMC studies, various techniques have been proposed to avoid such a bottleneck restricting the relaxation to the target distribution.

\section{Hamiltonian Monte Carlo}\label{sec:HMC}
We have seen that, in the simple gradient method, the state is updated in the direction along the gradient of the potential, which is normal to the energy surface. With such a method, it is difficult to avoid to be trapped in the local minimum of the potential. To overcome this difficulty, it has been proposed to add extra degrees of freedom to the original system to make new directions to escape from the local minimum of the potential. This idea is called an extended ensemble method. A method called Hamiltonian Monte Carlo~(HMC) is one of the realizations of the extended ensemble methods. In the HMC, in addition to the original state variable $x$, a momentum $p$ is introduced as an auxiliary variable. By introducing the momentum, the dimension of the dynamical system doubles, and it becomes easier to escape from the local minimum of the potential. In other words, when the kinetic energy exceeds the energy gap between the local minimum and the local maximum of the potential $U(x)$, the state can escape from the local minimum of the potential. The basic concept of the HMC is that the Gibbs distribution 
\begin{eqnarray}
\pi_{x, p}(x, p) &=& \exp\left[-H(x, p)/T\right] / Z_{x, p}\,,\label{xpGibbs}\\
H(x, p) &=& U(x) + \displaystyle\sum_i \dfrac{p_i^2}{2 m_i}
\end{eqnarray}
is invariant under the Hamiltonian dynamics 
\begin{eqnarray}
\dot{x}_i &=& \dfrac{p_i}{m_i}\,,\label{xHMC}\\
\dot{p}_i &=& -\dfrac{\partial U(x)}{\partial x_i}\,,\label{pHMC}
\end{eqnarray}
where $Z_{x, p}: = \int dx dp\, \exp\left[-H (x, p) / T \right]$ is a partition function. Here, $m_i$ represents the mass of the $i$-th degree of freedom. The target Gibbs distribution $\pi(x) = \exp\left[-U(x) / T  \right] / Z$ is acquired as a marginal distribution $\pi(x) = \int dp\, \pi_{x, p}(x, p)$ via the Gibbs distribution~(\ref{xpGibbs}). 

The algorithm of the HMC consists of the following steps. (i) Sample the momentum $p'_i$ ($i=1, \cdots, N$) from the Gaussian distribution
\begin{eqnarray}
P_{\rm G}(p'_i) = \dfrac{1}{\sqrt{2\pi m_i T}}\exp\left[-\dfrac{p_i^{\prime 2}}{2m_i T}\right]\,.\label{pGauss}
\end{eqnarray}
This procedure changes the state from $\left(x, p \right)$ to $\left(x, p'\right)$. (ii) Evolve the state for waiting time $\tau$ starting from the initial state $\left(x, p'\right)$ according to the Hamiltonian dynamics~(\ref{xHMC}) and (\ref{pHMC}). We denote the obtained state as $\left(x'', p''\right)$. (iii) According to the Metropolis-Hasting rule~\cite{Metropolis1953, Hastings1970}, the state obtained in the step (ii), $\left(x'', p''\right)$, is accepted with the acceptance rate $\min\left[1, \exp\left\{-\left[H(x'', p'') – H(x, p')\right] / T \right\} \right]$. Otherwise, the state remains at $\left(x, p'\right)$. The algorithm of the HMC consists of a repetition of these three steps.

Note that the Gibbs distribution~(\ref{xpGibbs}) is invariant under the Hamiltonian dynamics~(\ref{xHMC}) and (\ref{pHMC}). In particular, the Gaussian distribution~(\ref{pGauss}) gives the steady state distribution for the momentum. In step~(i), the momentum $p$ is sampled from this invariant distribution. The advantage of the HMC is that the Gaussian random variables can be easily generated in numerical manners. In step~(ii), the state update is ballistic on the energy surface. Even if the state is located at the local minimum of the potential $U(x)$, it is possible to escape from it by the effect of kinetic energy. The rejection in step~(iii) is exploited to eliminate nonphysical time evolution~\cite{hairer_lubich_wanner_2003}. Since the total energy is conserved under the Hamiltonian dynamics, the acceptance rate is theoretically always unity. However, naive numerical calculations have been reported to show an increase in total energy. The step~(iii) is introduced to eliminate this possibility to guarantee the calculation accuracy. Thus, step~(iii) is extra and can be omitted when the time evolution of the Hamiltonian dynamics is calculated with sufficiently high accuracy.

In the simple gradient method~(\ref{gradLangevin}), the state update in the normal direction of the energy surface is ballistic. The update on the energy surface is diffuse, since the state update on the energy surface is caused only by noise. On the other hand, in the HMC, the update in the normal direction of the energy surface is caused only by the random sampling of momentum. However, the update on the energy surface is ballistic since the state evolves according to the Hamiltonian dynamics. The Gibbs distribution obeys the principle of equal a priori weights for states with equal energy. The HMC is expected to quickly satisfy the principle of equal a priori weights by the ballistic state updates on the energy surface.

\section{Ohzeki-Ichiki method}\label{sec:OI}
The violation of the detailed balance condition was shown to accelerate relaxation to the steady state due to the eigenvalue shit for the Fokker-Planck operator~\cite{PhysRevE.88.020101}. In order to systematically introduce the violation of the detailed balance condition, Ohzeki and Ichiki have proposed to duplicate the original system to introduce a rotating probability current between the two duplicated systems: 
\begin{eqnarray}
dx_i(t) &=& -\dfrac{\partial U(x)}{\partial x_i}dt + \gamma\dfrac{\partial U(y)}{\partial y_i}dt + \sqrt{2T}dW_i^x(t)\,,\label{xOI}\\
dy_i(t) &=& -\dfrac{\partial U(y)}{\partial y_i}dt - \gamma\dfrac{\partial U(x)}{\partial x_i}dt + \sqrt{2T}dW_i^y(t)\,,\label{yOI}
\end{eqnarray}
where $x_i$ and $y_i$ are degrees of freedom belonging to the original and the replicated system, respectively. $W_i^x$ and $W_i^y$ are independent standard Wiener processes: 
\begin{eqnarray}
\left\langle dW_i^x(t) dW_j^x(t')\right\rangle &=& \delta_{ij}\delta(t-t')dt\,,\\
\left\langle dW_i^y(t) dW_j^y(t')\right\rangle &=& \delta_{ij}\delta(t-t')dt\,,\\
\left\langle dW_i^x(t) dW_j^y(t')\right\rangle &=& 0\,.
\end{eqnarray}
This system has the steady state distribution of Gibbsian form 
\begin{eqnarray}
\pi_{x, y}(x, y) = \exp\left\{-\beta\left[U(x)+U(y)\right]\right\}/Z_{x, y}\,,\label{xyGibbs}
\end{eqnarray}
where $\beta = 1/T$, and $Z_{x, y}$ is a partition function. 
Then, the target distribution $\pi(x) = \exp\left[-U(x)/T\right]/Z$ is acquired as the marginal distribution $\pi(x) = \int dy\, \pi_{x, y}(x, y)$. 
Note that this system violates the detailed balance condition, but satisfies the balance condition 
\begin{eqnarray}
\displaystyle\sum_{i}\dfrac{\partial}{\partial x_i}u_i^x \pi(x, y) + \sum_{i}\dfrac{\partial}{\partial y_i}u_i^y \pi(x, y)= 0\,,
\end{eqnarray}
where the driving force 
\begin{eqnarray}
u_i^x &=& \gamma\dfrac{\partial U(y)}{\partial y_i}\,,\\
u_i^y &=& -\gamma\dfrac{\partial U(x)}{\partial x_i}
\end{eqnarray}
yields the probability current characteristic to the violation of the detailed balance. The introduction of the driving force satisfying the balance condition remains the Gibbs distribution~(\ref{xyGibbs}) to be the steady state distribution. Although the two duplicated systems affect each other via the driving force, the steady state distribution for each system is independent.

In the Ohzeki-Ichiki dynamics~(\ref{xOI}) and (\ref{yOI}), the same form of the potential in the original $x$-system is chosen as that in the duplicated $y$-system. However, there is arbitrariness in the choice of the potential in the $y$-system, since $y$ is an auxiliary variable and the target distribution is given as the marginal distribution $\pi(x) = \int dy\,\pi_{x, y}(x, y)$. Therefore, the potential in the $y$-system does not have to be the same as that of the $x$-system. Consider the following dynamics: 
\begin{eqnarray}
dx_i(t) &=& \left[-\dfrac{\partial H_x(x)}{\partial x_i} + \gamma\dfrac{\partial H_y(y)}{\partial y_i}\right]dt + \sqrt{2T}dW_i^x(t)\,,\nonumber\\\label{xGOI}\\
dy_i(t) &=& \left[-\dfrac{\partial H_y(y)}{\partial y_i} - \gamma\dfrac{\partial H_x(x)}{\partial x_i}\right]dt + \sqrt{2T}dW_i^y(t)\,,\nonumber\\\label{yGOI}
\end{eqnarray}
where $H_x(x) = U(x)$ is the potential in the original $x$-system, and the energy $H_y(y)$ in the $y$-system can be in the form of an arbitrary function. This system has the following steady state distribution independent of the value of $\gamma$: 
\begin{eqnarray}
\pi_{x, y}(x, y) = \exp\left\{-\beta\left[H_x(x) + H_y(y)\right]\right\} / Z_{x, y}\,.
\end{eqnarray}
Therefore, the target distribution is obtained as a marginal distribution $\pi(x) = \int dy\, \pi_{x, y}(x, y)$ for an arbitrary form of $H_y$.

Consider the change of variables in dynamics~(\ref{xGOI}) and (\ref{yGOI}) as $\gamma = \tilde{\gamma}T$, $\tilde{t} = \tilde{\gamma}T t$. Then the dynamics 
\begin{eqnarray}
dx_i(\tilde{t})& =& \dfrac{\partial H_y(y)}{\partial y_i} dt\,,\label{limxHMC}\\
dy_i(\tilde{t})& =& -\dfrac{\partial H_x(x)}{\partial x_i} dt\label{limyHMC}
\end{eqnarray}
is obtained in the limit of $\tilde{\gamma}\to\infty$. 
Note that $H_x$ and $H_y$ play the roles of potential and kinetic energies in this dynamics, respectively. In fact, the choice of $H_y(y) = \sum_i y_i^2 / 2m_i$ reproduces the Hamiltonian dynamics~(\ref{xHMC}) and (\ref{pHMC}). In dynamics~(\ref{xGOI}) and (\ref{yGOI}), the driving force proportional to $\gamma$ causes the violation of the detailed balance condition. The case of $\gamma = 0$ corresponds to the simple gradient method. On the other hand, the dynamics in the limit $\gamma\to\infty$ corresponds to the Hamiltonian dynamics. Thus, it is concluded that the dynamics~(\ref{xGOI}) and (\ref{yGOI}) seamlessly connects the gradient method and the Hamiltonian dynamics that is the basis of the HMC.

\section{hybrid use of gradient method and Hamiltonian dynamics}\label{sec:hybrid}
In the previous section, we have introduced the dynamics, which incorporates the simple gradient method and the Hamiltonian dynamics. By the simple gradient method, the state update on the energy surface is realized diffusely, and it takes a long time to satisfy the principle of equal a priori weights. On the other hand, in the HMC, the state update on the energy surface is so ballistic that the principle of equal a priori weights is quickly satisfied. However, since the total energy is conserved under the Hamiltonian dynamics, transitions between energy surfaces are prohibited. For this reason, the HMC requires resampling of momentum from the Gaussian distribution~(\ref{pGauss}) which is realized in the steady state.

Consider the case of finite $\gamma$ in the dynamics~(\ref{xGOI}) and (\ref{yGOI}) with harmonic $H_y$ that connects the simple gradient method and the Hamiltonian dynamics. In such a dynamics, the state update on the energy surface, which has been a bottleneck of relaxation to the steady state in the simple gradient method, is realized to become ballistic. In addition, the effects of gradients and noise automatically enhance transitions between energy surfaces. Therefore, it is not required to resample the momentum, unlike the case of conventional HMC.

\begin{figure}
    \centering
    \includegraphics[width=\columnwidth]{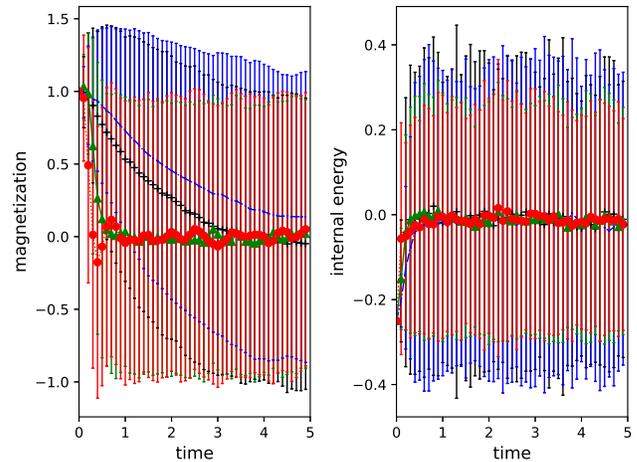}
    \caption{(Color online) Time evolution of the state $\left\langle x\right\rangle$ (left panel) and the internal energy (right panel). The black cross marks, blue dots, green triangles, and red circles indicate the results of the simple gradient method, conventional HMC, conventional Ohzeki-Ichiki method, and the proposed method, respectively. The error bars indicate variances.}
    \label{fig:fig1}
\end{figure}
To demonstrate the performance of our proposed method, i.e., the dynamics with harmonic $H_y$, we first deal with a toy model of a one-dimensional double-well potential: 
\begin{eqnarray}
U(x) = \dfrac{1}{4}x^4 - \frac{1}{2}x^2\,.
\end{eqnarray}
The initial condition is set to be in one of the potential wells at $x = 1$. Thus, the system must go beyond the potential hill at $x = 0$ to realize the steady state. In our numerical calculations, we set the temperature as $T = 1.0$. The infinitesimal time-step is set to be $dt = 1.0 \times 10^{-4}$. We compare the performance of the simple gradient method, the conventional HMC, the conventional Ohzeki-Ichiki method, namely, the dynamics~(\ref {xGOI}) and (\ref {yGOI}) with $H_x(x) = H_y(x) = U(x)$, and the proposed dynamics~(\ref {xGOI}) and (\ref{yGOI}) with $H_x(x) = U(x)$ and $H_y (y) = y^2 / 2m$. The time evolution of the Langevin equations is calculated by applying the Heun scheme~\cite{Kloeden1992}. The time evolution of the Hamiltonian dynamics in the HMC is calculated using the leapfrog method~\cite{hairer_lubich_wanner_2003}. Other parameters are set as follows: In the HMC, the particle mass is set as $m = 1$. In the algorithm of the HMC, it is necessary to evolve the Hamiltonian dynamics by a certain waiting time $\tau_{\rm wait}$ before resampling the momentum. We set the waiting time as $\tau_{\rm wait} = 0.01$. In the Ohzeki-Ichiki method, the parameter $\gamma$ characterizing the violation of the detailed balance condition is set as $\gamma = 10.0$. In the generalized Ohzeki-Ichiki method where $H_y(y)$ is harmonic, the particle mass is set as $m = 1.0$. The value of $\gamma = 10.0$ is also chosen in this dynamics. Figure.~\ref{fig:fig1} shows the numerical results averaged over $N_{\rm sample}=1000$ independent runs taking time average during $\Delta t = 0.1$. The Ohzeki-Ichiki method shows faster convergence to the steady state than the simple gradient method because of the detailed balance violation. Furthermore, it can be seen that the convergence of the proposed dynamics with harmonic potential for $y$ is faster than the Ohzeki-Ichiki dynamics, since the potential of the $y$-system is complicated in the conventional Ohzeki-Ichiki method. In the HMC, relaxation depends on the waiting time $\tau_{\rm wait}$. The larger $\tau_{\rm wait}$, the smaller the number of Monte Carlo steps is required for convergence. However, as seen in Fig.~\ref{fig:fig1}, it requires a longer calculation time, which is given by the product of $\tau_{\rm wait}$ and the Monte Carlo steps in HMC, than other methods. As seen in the previous section, the timescale conversion in the Ohzeki-Ichiki dynamics reproduces the Hamiltonian dynamics. Due to the limit of this timescale conversion, it is difficult to make a direct comparison between the HMC and the Ohzeki-Ichiki method. In fact, in the limit of $\tilde{\gamma}\to\infty$, $d\tilde{t}$ corresponding to the infinitesimal time step $dt$ diverges. This means that one Monte Carlo step in the Ohzeki-Ichiki method should be compared with the result of the HMC with the limit of long waiting time $\tau_{\rm wait}\to\infty$. 

We also evaluate the integrated auto-correlation time $\tau_{\rm int}: = \int_0^\infty dt'\left[\left\langle x (t) x(t + t') \right\rangle - \left\langle x \right\rangle^2 \right] / \left[\left\langle x^2 \right\rangle - \left\langle x \right\rangle^2 \right]$. The integrated auto-correlation time for each dynamics is evaluated by the empirical average after the convergence to the steady state. We obtain $\tau_{\rm int} = 2.00$ for the simple gradient method, which corresponds to the dynamics with $\gamma=0$, $\tau_{\rm int} = 0.19$ for the conventional Ohzeki-Ichiki method with $\gamma = 10.0$, and $\tau_{\rm int} = 0.14$ for the proposed hybrid use of the gradient method and the Hamiltonian dynamics with $\gamma=10.0$, respectively. In addition to the convergence of $\left\langle x\right\rangle$ and $\left\langle U(x)\right\rangle$ shown in Fig.~\ref{fig:fig1}, these results imply that the proposed method leads the significant reduction of the relaxation time to the steady state. 

To demonstrate the removal of the critical slowing down in our method, we next deal with the two-dimensional $XY$ model on a square lattice:
\begin{eqnarray}
U(x)=-\displaystyle\sum_{\langle i, j \rangle}\cos\left(x_i – x_j \right)\,,
\end{eqnarray}
where the sum is taken over all pairs of the nearest neighboring sites. The two-dimensional $XY$ model exhibits the Kosterlitz-Thouless transition at $T_{\rm c} = 0.89213(10)$~\cite{PhysRevB.52.4526}. At temperatures below $T_{\rm c}$, magnetization $m = \sum_{i=1}^N \sin x_i / N$ exhibits slow relaxation following the power law decay~\cite{Nishimori2010}. Since the critical slowing down is a bottleneck for convergence to the targeted steady state, it is preferred to avoid such slowing down behaviors. 

We compare the convergence performance of the gradient method, the Ohzeki-Ichiki method, and the proposed method, in which the potential of the $y$-system is given by $H_y(y) = \sum_{i=1}^N y_i^2 / 2m_i$. 
In our numerical calculations, the number of spins is set to be $N = 10 \times 10$. According to the finite size correction, the effective critical temperature for this system is evaluated as $T_{\rm ceff}\sim 0.975$~\cite{doi:10.1143/JPSJ.81.113001}. To demonstrate the removal of the critical slowing down, the temperature is set to be $T=0.5 < T_{\rm ceff}$. Other parameters are set as follows: the mass in the proposed dynamics is set as $m_i=1.0$ for each $i=1, \cdots, N$. The parameter $\gamma=5.0$ is chosen for both cases of $H_y(y)=U(y)$ and $H_y(y)=\sum_{i=1}^{N} y_i^2 / 2m_i$. The state with all spins in up-state, i.e., $x_i=\pi/2$ for all $i=1, \cdots, N$ is chosen as the initial state. The infinitesimal time step is set to be $dt=1.0\times 10^{-4}$. The Langevin equations are integrated by the Heun scheme. 

\begin{figure}
    \centering
    \includegraphics[width=\columnwidth]{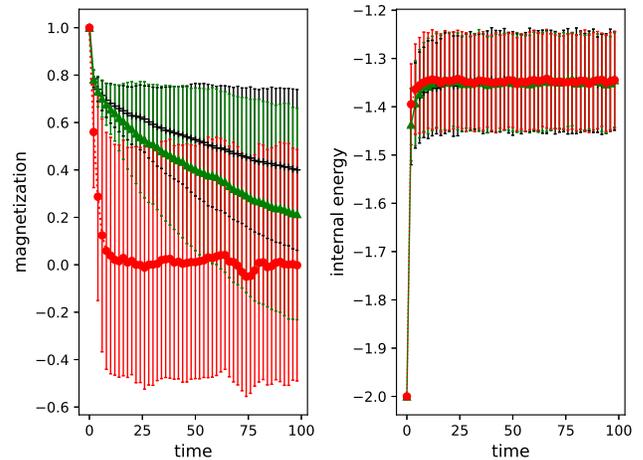}
    \caption{(Color online) Time evolution of the magnetization (left panel) and internal energy (right panel) of $XY$ model. The black cross marks, green triangles, and red circles indicate the results of the gradient method, conventional Ohzeki-Ichiki method, and the proposed method, respectively. The error bars indicate variances.}
    \label{fig:fig2}
\end{figure}

Figure~\ref{fig:fig2} shows the results averaged over $N_{\rm sample}=1000$ independent runs taking time average during $\Delta t = 0.1$. 
Although the simple gradient dynamics exhibits the critical slowing down, the Ohzeki-Ichiki and our proposed dynamics show faster convergence. 
The conventional Ohzeki-Ichiki method provides faster relaxation than the gradient method, and the significant improvement is achieved by the proposed method. In the proposed method, both magnetization and internal energy rapidly converge to the steady state values, and the critical slowing down appears to be eliminated.

\section{summary and discussion}\label{sec:summary}
We have seen that the Ohzeki-Ichiki method seamlessly connects the simple gradient method with the Hamiltonian dynamics. The Hamiltonian dynamics corresponds to a specific limit of the generalized Ohzeki-Ichiki method. 
The HMC does not satisfy the detailed balance condition in general. In the HMC, the candidate of the updated state depends on the waiting time, which defines a leapfrog operator $\hat{L}$. Even if the updated state $(x', p') = \hat{L}(x, p)$ is proposed starting from the current state $(x, p)$ by the leapfrog operator, the reverse transition $(x', p')\to (x, p)$ is not necessarily proposed. In other words, $(x, p) = \hat{L}(x', p')$ is not satisfied in general. The HMC with the detailed balance condition can be realized using a leapfrog operator $\hat{L}$ adaptively defined with an appropriate waiting time~\cite{MichikoOkudo2017}. In contrast, our proposed method does not constrain any leapfrog operator. The detailed balance condition is not satisfied, but the balance condition is in any timescale. 

The convergence performance of the HMC strongly depends on the waiting time. The method called No-U-Turn Sampler~(NUTS) adaptively determines the waiting time and efficiently proposes the updated state~\cite{JMLR:v15:hoffman14a}. This technique accelerates the convergence speed to the steady state with respect to Monte Carlo steps. However, in the HMC, a sufficient number of Monte Carlo steps are required for convergence, since the transition between energy surfaces occurs only when the momentum is resampled every single Monte Carlo step. Conversely, to shorten the simulation time, it is required to shorten the time of single Monte Carlo step, i.e., the waiting time. Note that the simulation time is proportional to the actual calculation time. Due to the trade-off relationship between the waiting time and Monte Carlo steps in the HMC, it takes a long simulation time to converge to the steady state. In contrast, our proposed method shows faster relaxation in simulation time. Therefore, it is concluded that our method shows better performance than the HMC with respect to actual calculation time.

It is worth mentioning that our method can be used in combination with other methods. For example, it can be used with the exchange Monte Carlo method and coarse-grained dynamics. In addition to this advantage, our proposed method can be easily applied to existing algorithms with detailed balance condition to improve their convergence speed. It is required only to add the driving force generating probability current and momentum dynamics to the original dynamics with the detailed balance condition. 

In the conventional Ohzeki-Ichiki method, the steady state distribution for the auxiliary variable $y$ has the same form as that for the original system. Thus, $y$ can be directly used to evaluate expectations for the target distribution. Note that the auxiliary variable $y$ doubles the number of samples in the procedure for evaluating the expectation in empirical manner. The obtained empirical average has less variance than that obtained by the dynamics with $H_y(y)\neq U(y)$. On the other hand, the choice of harmonic $ H_y $ exhibits faster convergence, but the auxiliary variable $y$ cannot be directly used to evaluate the expectations for the target distribution.

Since our method exploits the violation of the detailed balance condition, the probability current is generated in the system. The probability current realizes a biased sampling, resulting in accelerated convergence to the target distribution~\cite{PhysRevE.91.062105}. It is known in such a system that the convergence of the long-time average of physical quantities, namely, the empirical average, to the ensemble average is accelerated~\cite{Coghi2021}.

Note that the choice of $H_y$ still has some arbitrariness. In the conventional Ohzeki-Ichiki method, $H_y$ is chosen as the potential of the original system. In the method proposed in this paper, $H_y$ is chosen as a harmonic one, which is the bridge between the gradient method and the HMC. However, $H_y(y)$ can be a function of an arbitrary form. We have seen that $H_y(y) = U(y)$ and $H_y(y) = \sum_i y_i^2 / 2m_i$ show different convergence performance to the target distribution. The performance of the dynamics~(\ref{xGOI}) and (\ref{yGOI}) depends on the choice of $H_y$. Since the harmonic $H_y$ has only a single energy valley, it is expected to be relaxed quickly. Thus, the relaxation of the variable $x$ belonging to the original system is also expected to be accelerated. However, a detailed discussion of the optimal $H_y$ is a matter for the future. For example, it remains an open problem whether the optimal $H_y$ for convergence depends on $H_x$.

\begin{acknowledgments}
A. Ichiki was supported by JSPS KAKENHI Grants No. JP17H06469. 
\end{acknowledgments}

\bibliography{ref.bib}

\end{document}